\documentclass[prl, twocolumn, floatfix]{revtex4}
\setcitestyle{super}

\usepackage{graphicx}
\usepackage{hyperref}
\usepackage{amsmath}
\usepackage{color}
\usepackage{amssymb}

\begin{document}

\title{Optical pattern formation in self-focusing and self-defocusing diffractively thick media}
\author{G.\ Labeyrie$^{1}$\footnote{To whom correspondence should be addressed.}, I. Kre\v{s}i\'{c}$^{2,3}$, R.\ Kaiser$^{1}$, and T.\ Ackemann$^{4}$}
\affiliation{$^{1}$Universit\'{e} C\^{o}te d'Azur, CNRS, Institut de Physique de Nice, 06560 Valbonne, France}
\affiliation{$^{2}$Institute of Physics, Bijeni\v{c}ka Cesta 46, 10000 Zagreb, Croatia}
\affiliation{$^{3}$Institute for Theoretical Physics, Vienna University of Technology, Vienna 1040, Austria}
\affiliation{$^{4}$SUPA and Department of Physics, University of Strathclyde,Glasgow G4 0NG, Scotland, UK}

\begin{abstract}

Cold atomic clouds constitute highly resonant nonlinear optical media, whose refractive index can be easily tuned via the light frequency. When subjected to a retro-reflected laser beam and under appropriate conditions, the cloud undergoes spontaneous symmetry breaking and spatial patterns develop in the transverse cross-section of the beam. We investigate the impact of the sign of the light detuning from the atomic resonance on these patterns, thus directly comparing pattern formation for self-focusing and self-defocusing nonlinearities. Our observations emphasize the need for a ''diffractively thick'' medium description of the light-cloud interaction, where diffraction and nonlinear propagation inside the sample are taken into account.

\end{abstract}

\maketitle

\section{Introduction}
Since of few decades, cold and ultracold atomic ensembles have been extensively used as quantum simulators to investigate solid-state physics. One of the main advantages of these systems is their tunability, implemented through variations of an external parameter. For instance, an externally applied magnetic field can be used to adjust the magnitude and sign of contact interactions between atoms through the use of Feschbah resonances, a technique that has been essential to the investigation e.g. of the BEC-BCS crossover~\cite{Bourdel2004}.
 
Similarly the optical properties, either linear or nonlinear, of a cold atomic sample can be conveniently tuned by adjusting the laser frequency detuning from an optical line taking advantage of the sharp resonance. In the present work, we change the sign of the laser detuning to pass from a self-focusing nonlinearity (where the refractive index increases with laser intensity) to a self-defocusing one (where the refractive index decreases with laser intensity). We are then able to investigate the impact of this change on light-induced self-organization due to diffractive coupling~\cite{Firth1990}, without actually modifying the sample. Our observations emphasize the important role played by diffraction and nonlinear refraction inside the cloud in this experiment.

Spontaneous patterns formation in optical systems has been studied for three decades, using a variety of platforms such as hot atomic vapors~\cite{Grynberg1988, Grynberg1994, Ackemann1994}, liquid crystals~\cite{MacDonald1992, Ciaramella1993} and photorefractives~\cite{Honda1993, Schwab1999}. Cold atomic samples emerged in the years 2010 as interesting media where different kinds of optical nonlinearities can be used for pattern formation~\cite{Gauthier2011, Labeyrie2014, Ackemann2021}. In our group, we demonstrated self-organization based on spatial bunching of the cold atoms~\cite{Labeyrie2014}, saturation of the atomic transition~\cite{Camara2015}, and magnetism based on optical pumping~\cite{Labeyrie2018, Kresic2018, Kresic2019}. These various mechanisms can be selected in the \textit{same} atomic cloud by tuning external parameters such as the laser intensity and frequency detuning, or an external magnetic field. In the present work, we use the magnetic spin-based nonlinearity at zero magnetic field, which leads to the spontaneous appearance of an anti-ferromagnetic phase with a square symmetry~\cite{Labeyrie2018, Kresic2018}.

The first observations of transverse instability in hot~\cite{Grynberg1988} and cold~\cite{Gauthier2011} atomic vapors relied on counterpropagating beams interacting with the nonlinear medium. The analysis of this problem is challenging as diffraction and nonlinearity are acting simultaneously. Semi-analytical results were obtained for the length scales and threshold of the instability under simplified conditions~\cite{Grynberg1988, Firth1990b, Honda1993} but to our knowledge only one numerical simulation was ever performed due to the high demand on computational resources~\cite{Sandfuchs2001}. The system suggested in~\cite{Firth1990} is conceptually much simpler. The pump beam traverses a nonlinear medium which is thin enough so that diffraction is not important. Then diffraction takes places in vacuum (without nonlinearity) between the medium and a retro-reflecting feedback mirror (for details see below Section Experiment). Due to this separation between nonlinearity and diffraction, this system is easily tractable analytically and numerically, e.g.~\cite{Firth1990, D'Alessandro1991, D'Alessandro1992, Ackemann2021}. It is now natural to ask when this separability breaks down, i.e. reducing the feedback mirror distance, when one will see effects of diffraction also within the medium. This problem arises already in the context of linear optics as it is related to the transition between Raman-Nath diffraction from ''thin'' gratings and Bragg diffraction from ''thick'' gratings (e.g.~\cite{Pierce1973}) and in atom optics in the transition between Kapitza-Dirac and Bragg diffraction (e.g.~\cite{Schaff2014}). Hence it has also relevance in understanding the different regimes of superradiance in nonlinear atom optics~\cite{Schneble2003, Robb2005}. There is also a temporal analogue for the single mirror feedback system based on nonlinear fibre optics and dispersion replacing diffraction and ''diffraction'' within the nonlinear medium can be important~\cite{Kozyreff2006}. In a more general setting the interaction of Talbot-like self-imaging phenomena and nonlinearity has been discussed in harmonic generation~\cite{Zhang2010} and in nonlinear fiber optics~\cite{Zhang2014, Zajnulina2024} and is often referred to as nonlinear Talbot effect.

We first describe our experiment and the principles behind self-organization due to diffractive coupling. We then present the results of an experiment where the feedback distance is varied across positive and negative values, for two values of the laser detuning with opposite signs. We then discuss these results, based on both a thin medium model~\cite{Firth1990} and a more involved thick medium one~\cite{Firth2017}.

\section{Experiment}

\begin{figure}
\begin{center}
\includegraphics[width=1.0\columnwidth]{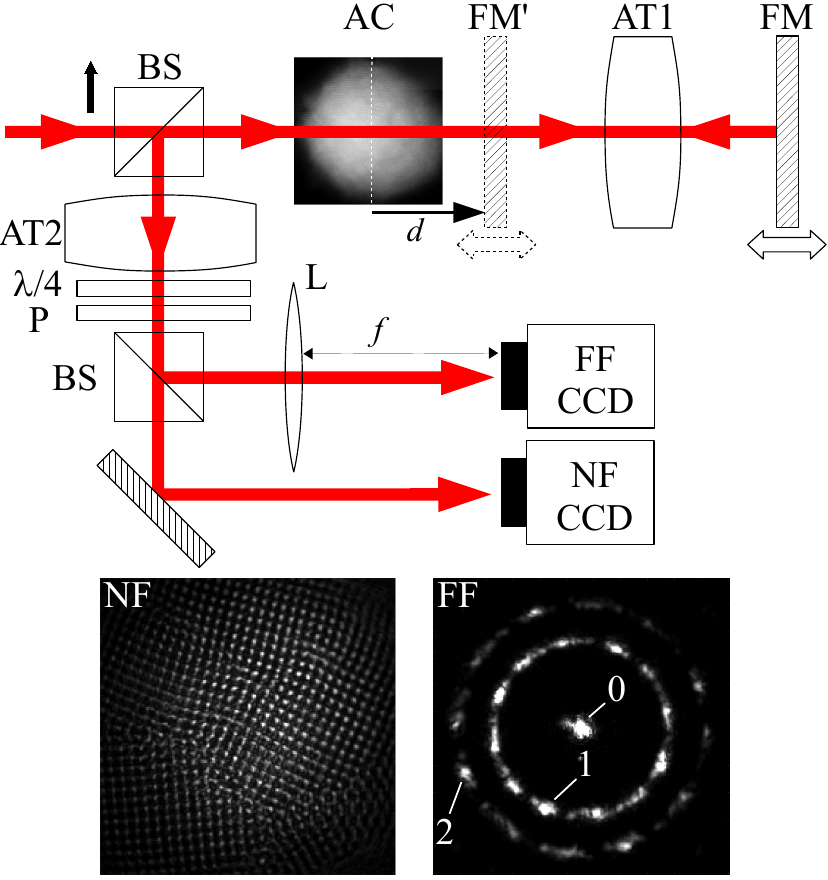}
\caption{Experimental setup. A laser beam is sent through a laser-cooled atomic cloud (AC) and retro-reflected by a feedback mirror (FM) on a translation stage. An afocal telescope (AT1) is used to image a ''virtual feedback mirror'' (FM') at a distance $d$ from the cloud. The reflected light is collected by a beam-splitter (BS), passed through a circular polarization analyzer (quarter-waveplate $\lambda/4$ + polarizer P), divided in two parts by a beam-splitter(BS) and directed on two CCDs used to record the transverse intensity distributions of the light in both near-field (NF) and far-field (FF ; 0: on-axis pump, 1: first off-axis wavenumber, 2: second off-axis wavenumber).}
\label{fig1}
\end{center}
\end{figure}

We use a modified version of our previous setup~\cite{Labeyrie2014, Camara2015, Labeyrie2018} sketched in Fig.~\ref{fig1}, based on the single feedback mirror scheme introduced by Firth in 1990~\cite{Firth1990}. A large cloud of cold $^{87}$Rb (full width at half maximum (FWHM) $L = 12$ mm, optical density at resonance = 130, temperature = 200 $\mu$K) is produced in a vapor-loaded magneto-optical trap (MOT). We then shut down the MOT and send a ''pump'' laser beam of waist 2.2 mm through the cloud. This beam is linearly polarized, with a detuning (either positive or negative) $\left|\delta\right| = 10.4~\Gamma$ from the $F = 2 \rightarrow F^{\prime} = 3$ transition, where $\Gamma$ is the natural width. The peak intensity is $I_0 = 13$ mW/cm$^2$, which corresponds to a saturation parameter $s = 0.008$ (assuming a saturation intensity of 3.58 mW/cm$^2$ for unpolarized atoms). The beam is applied for a duration of 1 ms, the detection (image acquisition) taking place in the last $100~\mu$s of this pulse. The beam transmitted by the cloud (corresponding to 75$\%$ of the incident) is retro-reflected by a ''feedback mirror'' set on a translation stage, located outside the vacuum chamber. An afocal telescope AT1 is used to create a ''virtual'' mirror near the cloud (i.e. inside the vacuum chamber) by imaging the real one. This arrangement allows us to have access to a small effective cloud-mirror distance $d$, as well as negative values~\cite{Ciaramella1993} (see Fig.~\ref{fig4}). As will be seen, the sign of $d$ plays an important role in the observations. After passing again through the cloud, part of the retro-reflected beam is collected by a non-polarizing beam-splitter for pattern detection. We select circularly-polarized light with the help of a quarter-wave plate and a polarizer. We then split this light in two parts that are directed to two CCD cameras to detect both the near-field (NF) and far-field (FF) intensity distributions simultaneously. The NF imaging is achieved through another afocal telescope AT2, the FF imaging requiring an additional lens L. During the interaction of the laser beam with the atoms, the magnetic field is set to $B = 0$.

\begin{figure}
\begin{center}
\includegraphics[width=1.0\columnwidth]{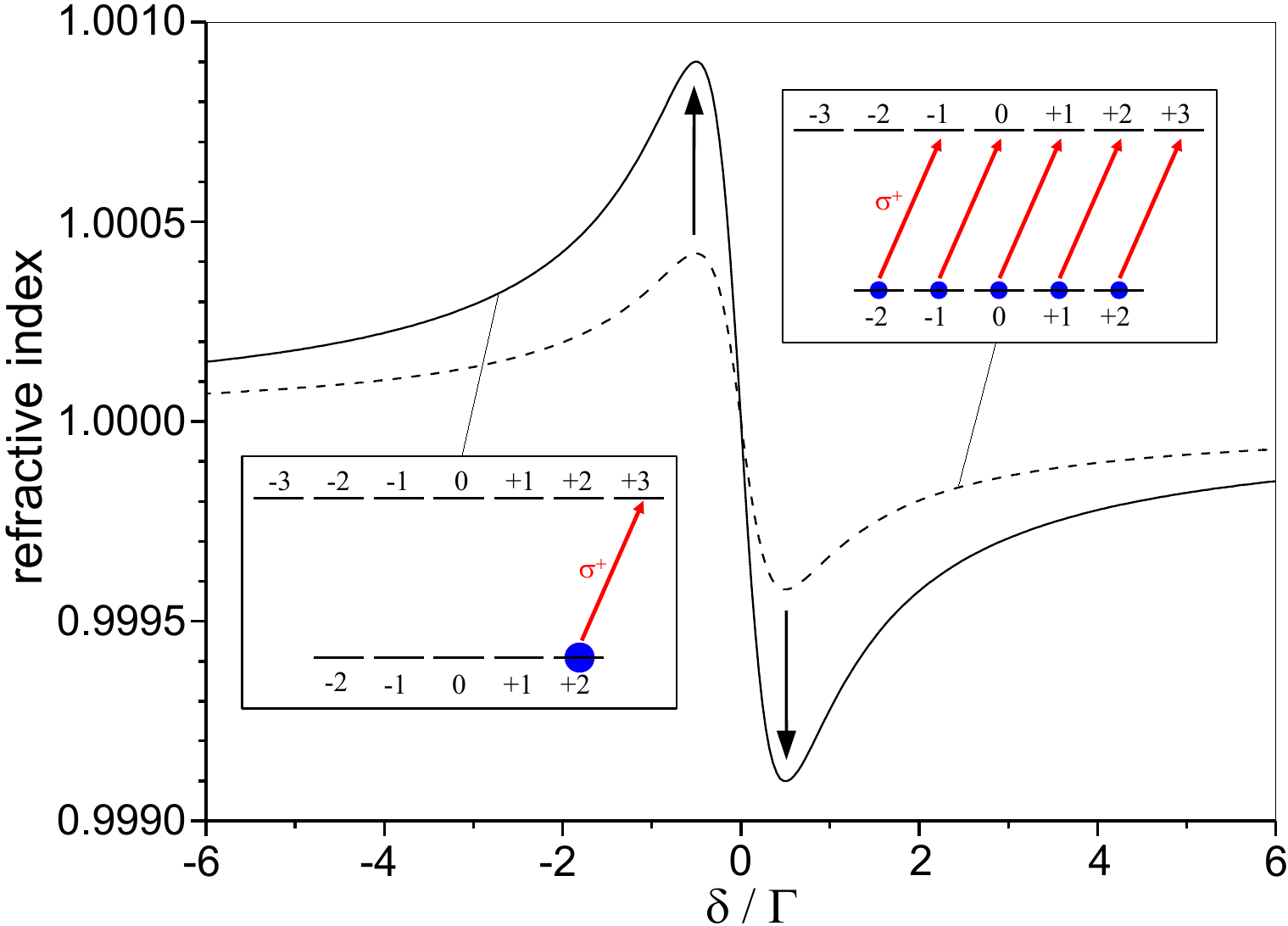}
\caption{Nature of the optical pumping nonlinearity. We plot the refractive index of the cold cloud versus light detuning for an atomic density $\rho = 10^{11}$ cm$^{-3}$ and circular polarization of the light. The dashed line corresponds to atoms evenly distributed among Zeeman sublevels, corresponding to a degeneracy factor $g = 7/15$. The solid line correspond to atoms fully pumped in the streched state. This can be obtained by increasing slightly the light intensity $I$ to overcome e.g. the redistribution effect of stray magnetic fields. Thus, the refractive index increases with $I$ for $\delta < 0$ (self-focusing nonlinearity) and decreases when $I$ increases for $\delta > 0$ (self-defocusing nonlinearity), as shown by the arrows.}
\label{fig2}
\end{center}
\end{figure}

Most measured quantities presented in this paper are extracted from far-field images (i.e. optical Fourier spectra) such as shown in Fig.~\ref{fig1}. The central spot corresponds to the pump beam. Above the instability threshold, it is surrounded by bright dots distributed on several circles corresponding to different transverse modes generated spontaneously. In the following we will refer to the collection of spots on a given circle as a ''mode'', as we are only interested on the transverse wavenumber. Two such modes are visible in the FF image in Fig.~\ref{fig1}, labeled 1 and 2. The radius of each circle corresponds to a diffracted angle $\theta_d = q / k = \lambda / \Lambda$, where $q = 2 \pi/\Lambda$ is the transverse wavenumber and $\Lambda$ the modulation wavelength in the transverse plane (determining the size of the pattern's unit cell in NF). We see in the NF image of Fig.~\ref{fig1} that the patterns have a clear square symmetry, but that the beam's cross-section is divided into several domains with different pattern orientation (''polycrystalline pattern''). Hence, the corresponding FF image exhibits here twelve spots (three sets of four spots). To perform the FF analysis, we first average ten single-shot FF images. Since both translational and rotational degrees of freedom are almost unconstrained in the patterns (the Gaussian shape of the input laser beam only providing a weak pinning of the position), the position and orientation of the patterns vary from shot to shot. As a result, the averaged FF image shows an intensity distributed more or less uniformly along the circle of radius $\theta_d$. By measuring $\theta_d$ we obtain $q$ and thus $\Lambda$. We also extract the diffracted power $P_d$, obtained by integrating the FF image around the circle. This quantity provides an indication of the distance to the instability threshold (the larger $P_d$, the larger the distance to threshold), although it is not as direct as measuring the threshold intensity which is more time-consuming. 

\begin{figure}
\begin{center}
\includegraphics[width=1.0\columnwidth]{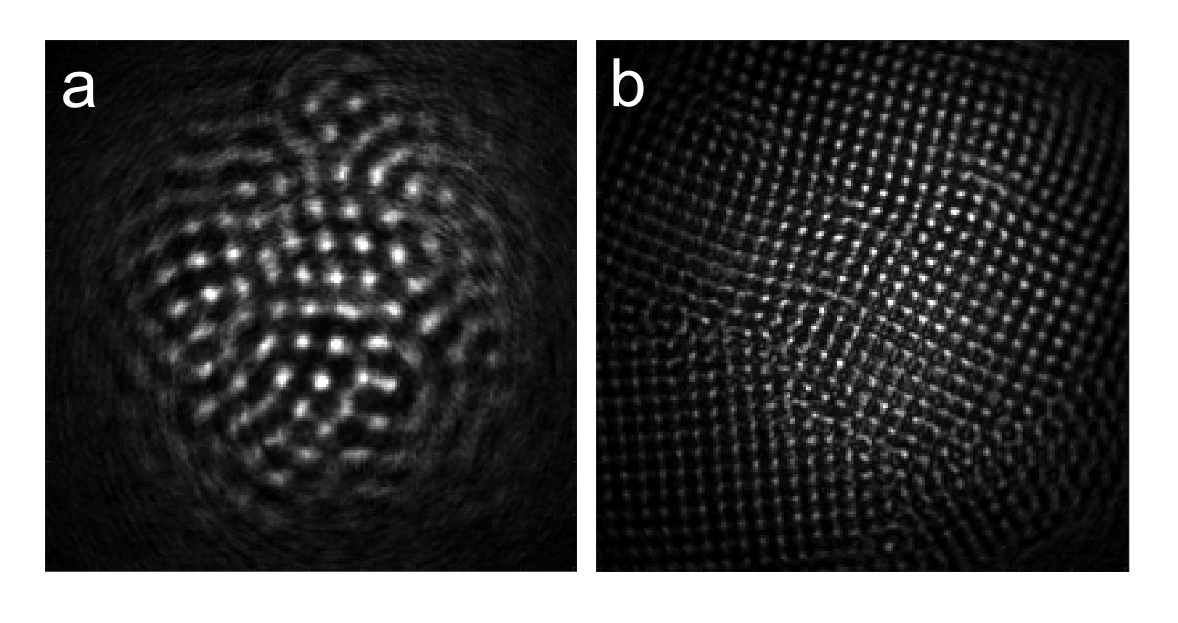}
\caption{Comparison of red and blue patterns (near-field). (a) NF image for $\delta = +10.4~\Gamma$. (b) NF image for $\delta = -10.4~\Gamma$. Both images are recorded in the $\sigma^+$ channel. The field of view is $3.6$ mm. Parameters: $d = -24.2$ mm, $I_0 =~$13 mW/cm$^2$, $B = 0$.}
\label{fig3}
\end{center}
\end{figure}

As mentioned earlier, the present work is based on the magnetic optical pumping nonlinearity~\cite{Labeyrie2018, Kresic2018, Kresic2019} where the patterns are due to a spatial modulation of the Zeeman populations in the ground-state (orientation). These patterns are very sensitive to the magnetic field, which needs to be controlled within the 10 mG range. When the magnetic field is set to zero, the system self-organizes into an anti-ferromagnetic phase where atoms are optically pumped in the stretched states. Because the Clebsh-Gordan coefficients are maximal in these stretched states, this optical pumping nonlinearity is self-focusing, i.e. the refractive index increases with light intensity, for red detunings ($\delta < 0$) and self-defocusing for blue ones ($\delta > 0$). This is illustrated in Fig.~\ref{fig2}, where we plot the (linear) refractive index for a cold cloud with an even population distribution in the Zeeman substates of the ground state (dashed line), and the same for a cloud optically pumped in the stretched state $m_F = +2$ (solid line). The optical pumping results in an increase of the refractive index for $\delta < 0$ and a decrease for $\delta > 0$. Note that this is opposite to the case of the standard 2-level nonlinearity~\cite{Camara2015}. Since the nonlinearity used in this work is based on the very efficient process of optical pumping, the associated intensity instability threshold is very low compared to that of instabilities based on other nonlinear mechanisms such as saturation~\cite{Camara2015} and opto-mechanics~\cite{Labeyrie2014}. Thus, the width of the dispersion curve for the cloud's refractive index is close to the natural width ($\Gamma / 2\pi \approx 6$ MHz), and one can conveniently pass from a self-focusing nonlinearity to a self-defocusing one be tuning the frequency of the light by a few tens of MHz, without modifying the size or geometry of the cloud.

The expected impact of the sign of the detuning on self-organization is twofold. Firstly, it affects the spatial period of the patterns. Indeed, in the thin medium model of diffractive coupling providing the optical feedback~\cite{Firth1990}, diffraction within the medium is neglected and only the propagation of light in free space between sample and mirror and back of length $2d$ converts a pure phase modulation into a pure amplitude modulation. A positive feedback is obtained after a round-trip propagation of $Z_T/4$ in the self-focusing case and $3Z_T/4$ in the self-defocusing case~\cite{Ackemann2021}, where $Z_T = 2 \Lambda^2 / \lambda$ is the so-called Talbot length.
These conditions for positive feedback, necessary for the establishment of the so-called ''Talbot modes'', can be summarized as follows. For a self-focusing nonlinearity (here for $\delta < 0$):

\begin{equation}
d = (1/4 + n).\frac{\Lambda^2}{\lambda}
\end{equation}
with n integer (positive or negative depending on the sign of $d$).
For a self-defocusing nonlinearity ($\delta > 0$):

\begin{equation}
d = (3/4 + n).\frac{\Lambda^2}{\lambda}
\end{equation}

The smallest-$q$ (largest $\Lambda$) Talbot mode for a self-focusing nonlinearity is thus characterized by a diffraction angle:

\begin{equation}
\theta_d = \sqrt{\frac{\lambda}{4d}},~~~~~d > 0
\end{equation}

and
\begin{equation}
\theta_d = \sqrt{-\frac{3\lambda}{4d}},~~~~~d < 0
\end{equation}

The reverse applies for a self-defocusing nonlinearity:

\begin{equation}
\theta_d = \sqrt{\frac{3\lambda}{4d}},~~~~~d > 0
\end{equation}

and
\begin{equation}
\theta_d = \sqrt{-\frac{\lambda}{4d}},~~~~~d < 0
\end{equation} 

Secondly, the sign of the detuning may affect the stability of the patterns due to nonlinear propagation effects taking place inside the cold atoms cloud. Indeed, our system can not be expected to behave according to the thin medium regime when $|d|$ is of the order of the thickness $L$ of the cloud. On the contrary, in this thick medium regime effects such as diffraction and nonlinear refraction take place within the cloud. Indeed, the bright spots seen in Fig.~\ref{fig3} can be individually seen as Gaussian beams with a waist of the order of a few tens of microns. Self-focusing or defocusing can then occur within the thickness of the cloud. Self-focusing is expected to balance diffraction and help stabilize the transverse size of the individual spots, while the opposite should occur for the self-defocusing situation~\cite{Labeyrie2011}. We thus expect instability threshold intensities for $\delta > 0$ to be substantially higher than for $\delta < 0$.

\section{Results}
Fig.~\ref{fig3} shows examples of NF images of patterns obtained for opposite values of the detuning ($\left|\delta\right| = 10.4~\Gamma$) and a negative feedback distance $d = -24.2$ mm. Since the optical thickness of the cloud is symmetrical with respect to the sign of $\delta$, the only change between the two situations is due to the refractive index which is dispersive (see Fig.~\ref{fig2}). As can be seen, the spatial scale $\Lambda$ of the patterns is larger for $\delta > 0$ than for $\delta < 0$. According to the discussion above, this observation confirms that the nonlinearity is self-focusing for $\delta < 0$ and self-defocusing for $\delta > 0$.

\begin{figure}
\begin{center}
\includegraphics[width=1.0\columnwidth]{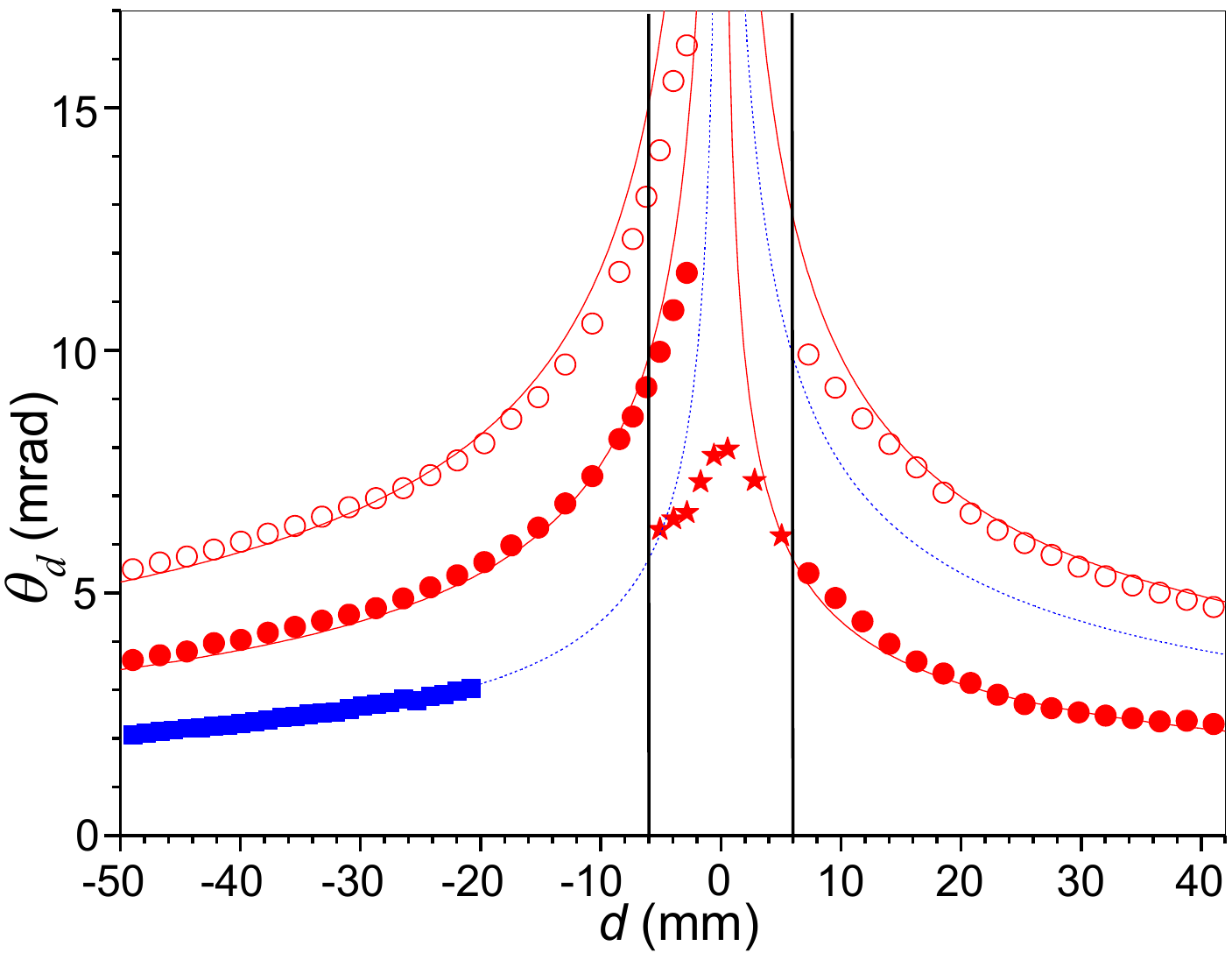}
\caption{Diffraction angle $\theta_d$ versus feedback mirror distance $d$. We report experimental measurements for the lowest-$q$ modes for $\delta < 0$ (circles) and $\delta > 0$ (squares). The solid red and dashed blue lines correspond to the thin medium model predictions for self-focusing and self-defocusing nonlinearities respectively. When the effective mirror is inside the cloud (delimited by the vertical bold lines), we observe strong deviations from the thin medium model ($\delta < 0$, stars).}
\label{fig4}
\end{center}
\end{figure}

A more quantitative comparison is provided in Fig.~\ref{fig4}, where we report the measured diffraction angle $\theta_d$ versus feedback distance $d$, for both $\delta < 0$ (red circles, dots and stars) and $\delta > 0$ (blue squares). For comparison, we also plot the thin medium predictions for a self-focusing nonlinearity (red solid lines) and a self-defocusing nonlinearity (blue dashed lines), as obtained from Eqs. (3)-(6). The vertical bold lines in this figure delimit the FWHM extent of the cloud, but it should be stressed that the spatial density profile is not uniform but close to a Gaussian. For clarity, we restrict the data presented in this figure to the two lowest-$q$ modes for $\delta < 0$ (except in the narrow range -6 mm $< d < -4$ mm where three modes are shown), and to the lowest-$q$ mode for $\delta > 0$. Note however that we can observe up to eight modes for $\delta < 0$ and large values of $\left|d\right|$, but only two modes for $\delta > 0$ (the second one much weaker).

Looking first at the $\delta < 0$ case, we see that when $\left|d\right|$ is large enough the measured $\theta_d$ closely matches the thin medium prediction. Note that the only adjustable parameter for this experiment-model comparison is the position $d = 0$, which is difficult to measure accurately in the experiment because of the large size of the cloud. A small global horizontal translation of the experimental curves is thus allowed to optimize the match.

When $\left|d\right|$ decreases below a certain value (which increases with the order of the considered mode) the measured $\theta_d$ starts to deviate from thin medium theory. When the effective feedback mirror is inside the cloud, this deviation becomes very important (see stars in Fig.~\ref{fig4}) and $\theta_d$ tends to saturate to a finite value for $d \approx 0$. This behavior was discussed in Ref.~\cite{D'Alessandro1992}: when $\left|d\right| \gg L$ the pattern's period $\Lambda$ is determined by feedback from the mirror and thus varies with $d$, whilst for $\left|d\right| < L$ it is the thickness of the sample that is the relevant length scale as in the situation with independent counter-propagating beams~\cite{Grynberg1988, Firth1990b} and $\Lambda = \sqrt{2 L \lambda}$. Using the cloud's FWHM size $L = 12$ mm, we get $\theta_d = 5.7$ mrad which is not too far from the saturation value observed in Fig.~\ref{fig4}. Of course, a quantitative difference is expected between the homogenous slab geometry and that of a Gaussian density profile. Looking in detail at Fig.~\ref{fig4} around $d = 0$, we observe a rather complex behavior with a smooth transition from thin medium behavior to a progressive saturation on the $d > 0$ side, and an abrupt mode jump occurring around $d = -6$ mm. We will discuss these observations later (see Fig.~\ref{fig6}).    

For $\delta > 0$ the situation is markedly different. The lowest-$q$ mode is observed only for large negative feedback distances $d < -20$ mm. No patterns are observed for $d > 0$ (see below for a discussion of this striking asymmetry). For $d < -20$ mm, the agreement with the thin medium theory is excellent confirming the self-defocusing nature of the nonlinearity for $\delta > 0$.

\begin{figure}
\begin{center}
\includegraphics[width=1.0\columnwidth]{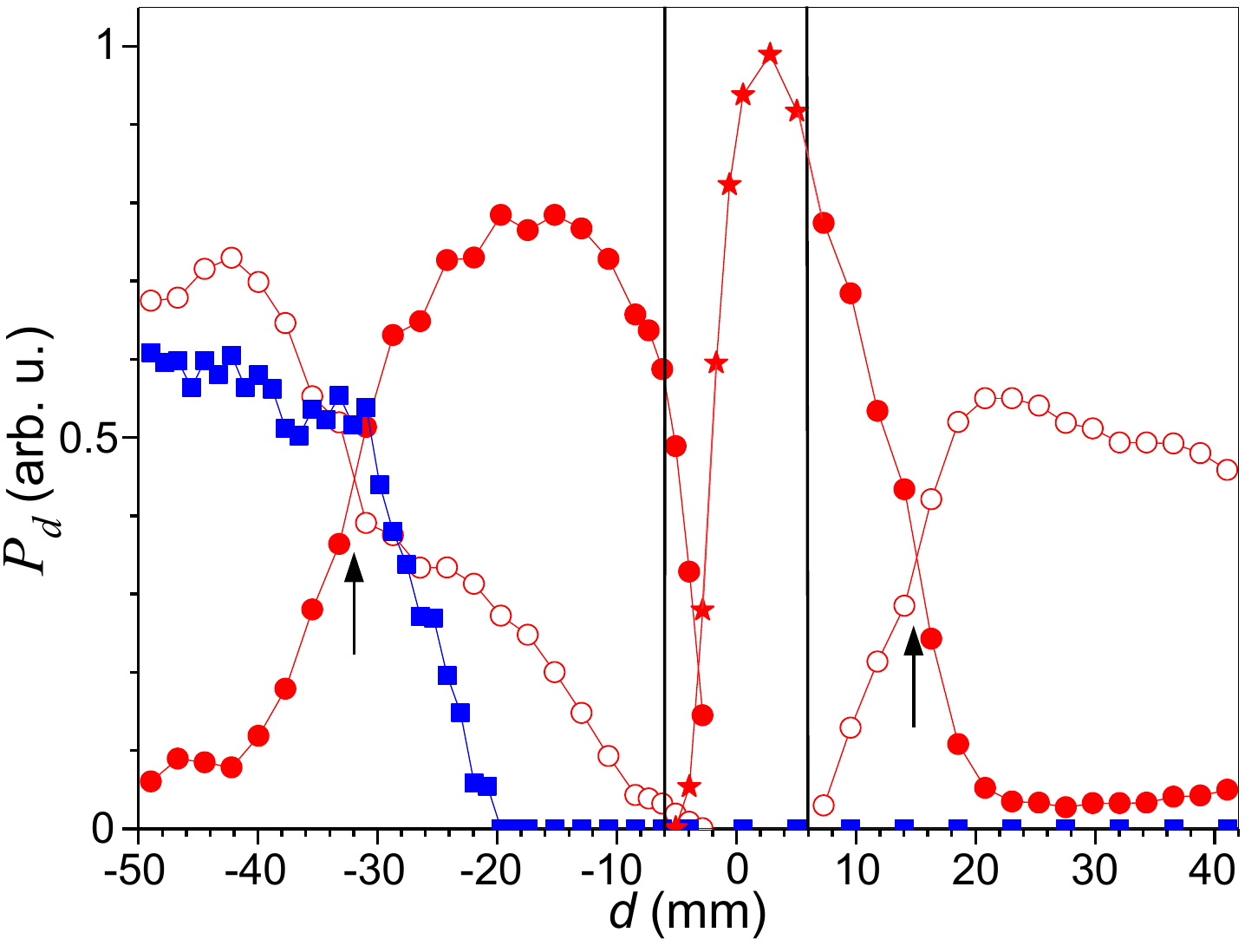}
\caption{Measured diffracted power $P_d$ versus feedback mirror distance for the data of Fig.~\ref{fig4}. The choice of symbols is the same as in Fig.~\ref{fig4}. The bold vertical lines delimitate the FWHM extent of the cloud. The arrows mark the positions of cross-over between the two modes for $\delta < 0$.}
\label{fig5}
\end{center}
\end{figure}

In complement to the data of Fig.~\ref{fig4}, we plot in Fig.~\ref{fig5} the diffracted power $P_d$ versus $d$ with the same choice of symbols. For $\delta < 0$, we observe the following general behavior for the Talbot modes (red dots and circles) including the higher-$q$ modes not shown here : the curves $P_d(\left|d\right|)$ are bell-shaped, and reach their maxima for values of $\left|d\right|$ that increase with the order of the Talbot modes. As a consequence, when increasing $\left|d\right|$ from zero we observe that the order of the dominant mode (i.e that with largest $P_d$) increases. The transition between the first and second Talbot modes occurs in Fig.~\ref{fig5} for $d = -32$ mm and $d = +15$ mm (arrows), so there is an asymmetry between $d > 0$ and $d < 0$. This behavior of transition between successive dominant modes as $\left|d\right|$ is increased was already observed in another experiment using the same optical pumping nonlinearity but with different parameters (see Fig. 15b of~\cite{Firth2017}).

\section{Discussion}       

The behaviors observed in Fig.~\ref{fig5} (and in Fig.~\ref{fig4} for small enough values of $\left|d\right|$) cannot be understood using the thin medium theory. In Ref.~\cite{Firth2017}, we described a thick medium model based on a 2-level description of the atom-light interaction (''saturable Kerr model''). This model in its most complete form included nonlinear absorption and propagation effects as well as diffraction inside the sample, and the presence of transverse and longitudinal gratings due to the interference between all light beams. In the simplified version used in this paper, we implemented a third-order expansion of the nonlinearity neglecting absorption to obtain a ''Kerr model'' where the nonlinearity is proportional to the intensity, without saturation. In this regime, it is possible to derive an expression for the instability intensity threshold. The details of this derivation are exposed in the Appendix. We stress that since the nonlinearity at work in this paper is based on Zeeman optical pumping instead of the saturation of a 2-level atomic transition, we do not expect a quantitative agreement between theory and experiment for instance on the threshold intensities.

\begin{figure}
\begin{center}
\includegraphics[width=1.0\columnwidth]{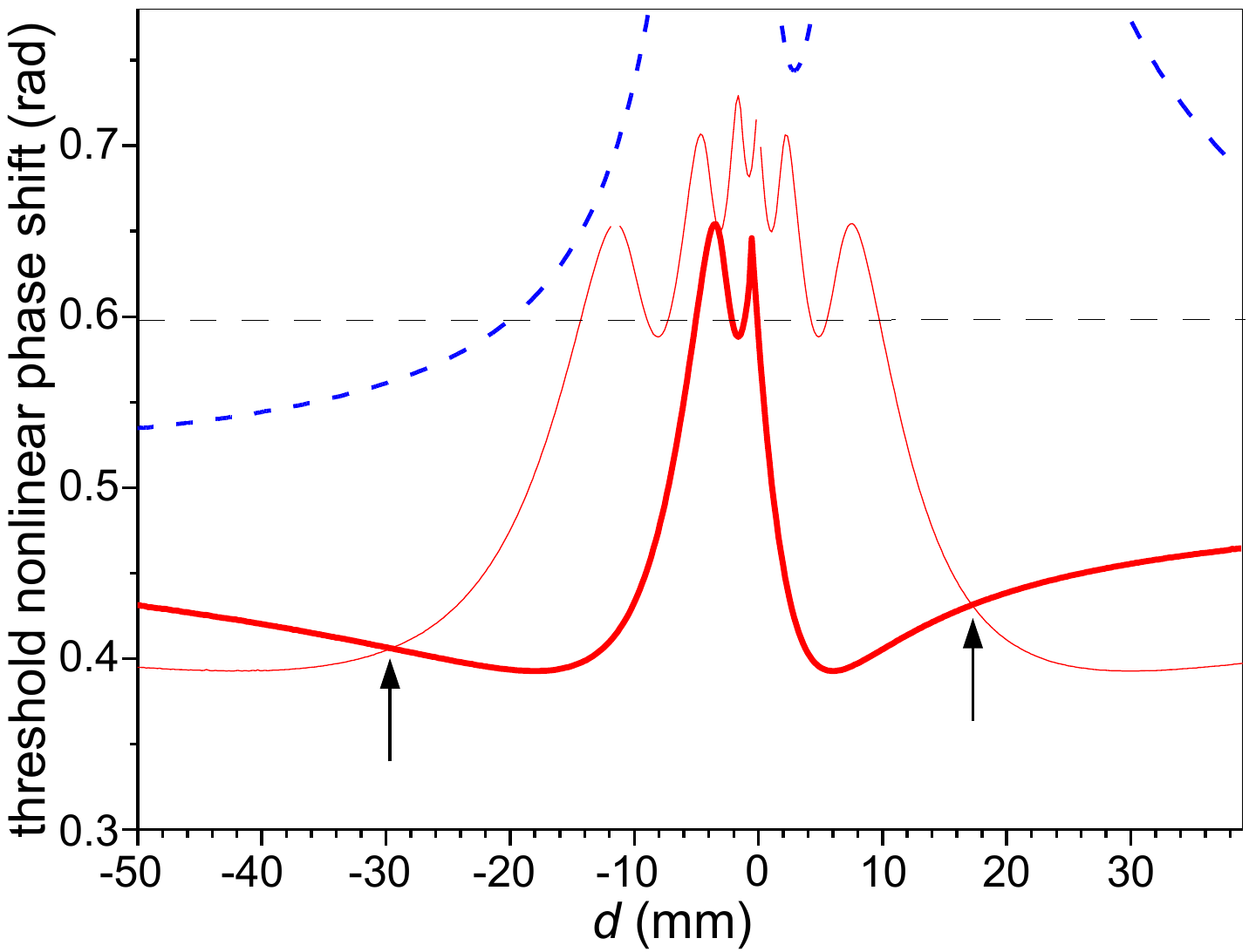}
\caption{Threshold nonlinear phase shift versus feedback mirror distance as predicted by the thick medium model. The bold and thin red curves correspond to the first and second lowest-$q$ modes for $\delta < 0$ respectively. The arrows mark the positions of cross-over between these two modes. The dashed blue curve correspond to the lowest-$q$ mode for $\delta > 0$. The horizontal dashed line marks the approximate value of the nonlinear phase shift in the experiment (see text).}
\label{fig6}
\end{center}
\end{figure}

In the framework of this model, a complex oscillatory behavior of the instability threshold with both the pattern's transverse wave vector $q$ and the feedback distance $d$ is predicted, causing the observed alternation of dominant Talbot modes as $d$ is scanned. On the contrary, in the case of self-defocusing, the thick medium model predicts a continuous decrease of the Talbot modes thresholds as $\left|d\right|$ increases, which is consistent with the observations reported in Fig.~\ref{fig5} for $\delta > 0$.

As already stressed above, a striking result of our experiment is the absence of patterns for $\delta > 0$ and $d > 0$. To try to understand this observation, we compare in Fig.~\ref{fig6} the intensity thresholds curves versus $d$, derived from our thick medium model (see the Appendix for details) for both $\delta < 0$ and $\delta > 0$. We plot in Fig.~\ref{fig6} the nonlinear phase shift at the instability threshold, which is proportional to the threshold intensity. The bold and thin red curves correspond to the first an second lowest-$q$ modes for $\delta < 0$ respectively. The dashed blue curve corresponds to the lowest-$q$ mode for $\delta > 0$. 

It is important to realize that the thin medium theory for a local Kerr medium as considered here predicts a threshold nonlinear phase shift of 0.5, independent of wavenumber and hence of $d$. Looking at Fig.~\ref{fig6}, we see that the thick medium threshold for the lowest-$q$ mode is considerably lower than 0.5 for the focusing medium (except for small $\left|d\right|$), and larger for the defocusing medium. This makes sense as self-focusing will stabilize ''spots'' to tubes along the whole longitudinal extent of the medium, whereas self-defocusing will destabilize these tubes.   

For $\delta < 0$, we observe a cross-over between the two lowest-$q$ modes for $d \approx -30$ mm and $d \approx +17$ mm (arrows in Fig.~\ref{fig6}), the lowest-$q$ mode being dominant for $-30$ mm $< d < 17$ mm (lowest intensity threshold). This is in excellent agreement with the experimental observation of Fig.~\ref{fig5}. We note that the asymmetry between $d < 0$ and $d > 0$ is well reproduced by the model.

We now turn to the even stronger asymmetry observed for $\delta > 0$. The fixed intensity $I_0$ used in the experiment corresponds to an horizontal line in Fig.~\ref{fig6}. Whenever the threshold intensity for a given mode is below  $I_0$, this mode is observed. As stressed before, we don't have a quantitative model to link the nonlinear phase shift to the laser intensity in the case of the magnetic optical pumping nonlinearity. We thus position the horizontal dashed line in Fig.~\ref{fig6} to match the region where patterns are observed for $\delta > 0$ ($d < -20$ mm). Once this is set, we see that patterns for $\delta > 0$ cannot be observed in the rest of the explored $d$ range, while for $\delta < 0$ patterns are observed throughout most of the $d$ range (with some oscillations around $d = 0$). The thick medium model thus provides an accurate description of our experimental observations. Extrapolating on the dashed blue curve in Fig.~\ref{fig6}, one would expect to be able to observe patterns for $\delta > 0$ at the same laser intensity but for a larger positive $d$ (not feasible with the actual setup), or by setting e.g. $d = +40$ mm and increasing $I_0$. We tried the latter, without success. We thus conclude that some additional effect, not included in our thick medium model, further prevents pattern formation for $\delta > 0$ and $d > 0$. We also measured the threshold intensities for $\delta > 0$ and $\delta < 0$ for $d = -50$ mm, and found that the threshold intensity for $\delta > 0$ is twice larger than for $\delta < 0$. This ratio is quite larger than what is reported in Fig.~\ref{fig6}. The likely effect missing in the current description is the residual motion of the atoms, leading to an increase of threshold with increasing $q$. The interplay of this effect with the self-defocusing in the medium is likely to suppress the instability. However, its modeling should be quite involved and is outside the scope of this paper.

\begin{figure}
\begin{center}
\includegraphics[width=1.0\columnwidth]{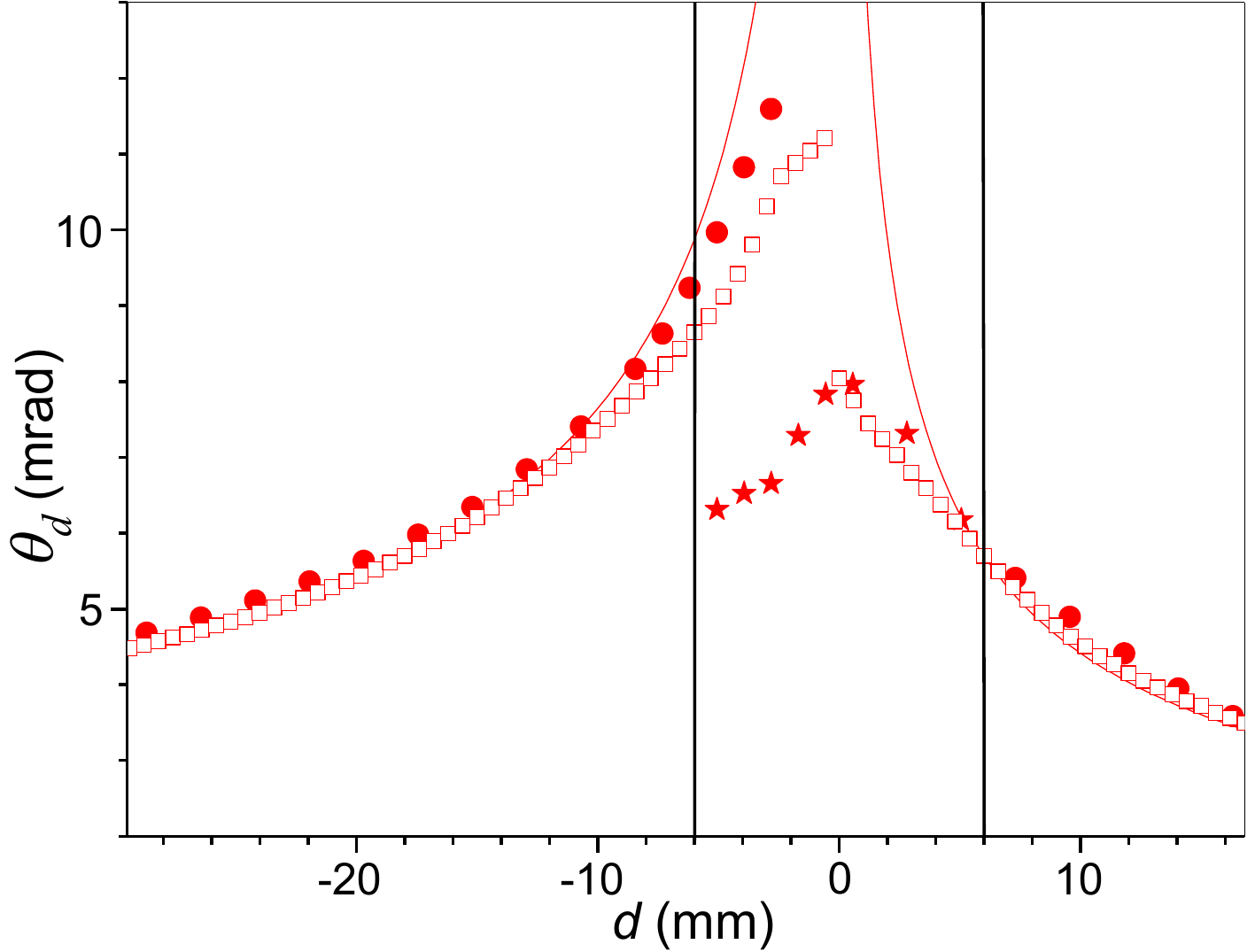}
\caption{Comparison between experiment and thick medium model. We show the data of Fig.~\ref{fig4} for $\delta < 0$ (lowest-$q$ mode, solid dots and stars) and a restricted range around $d = 0$. The red solid lines correspond to the thin medium model. The squares are obtained from the thick medium model.}
\label{fig7}
\end{center}
\end{figure}

Finally, we compare in Fig.~\ref{fig7} the experimentally-measured $\theta_d$ to the thick medium prediction. The dots and stars correspond to the experimental data for the lowest-$q$ mode as in Fig.~\ref{fig4}, the red curves showing the thin medium theory. The open squares are obtained with the thick medium model. Whilst the agreement is not perfect, the main features (saturation and mode jump) are well reproduced.  

\section{Conclusion}
In this work, we used the resonant character of the nonlinear refractive index of a cold atomic cloud to directly compare optical pattern formation for self-focusing and self-defocusing nonlinearities in the single feedback mirror configuration. We find that the pattern's respective sizes in the two situations are in quantitative agreement with predictions of the thin medium model when the feedback distance $d$ is large compared to the cloud's size $L$. However, even in this range of feedback distances, thick medium effects manifest themselves through the $d$-dependence of threshold intensities for the different Talbot modes. For $\delta > 0$, the reduced range of observation of the patterns is also attributed to thick medium effects, namely self-defocusing inside the cloud. 

We also reported the first detailed experimental observation of the behavior of the pattern size around $d = 0$ in a diffractively thick medium. When the effective feedback mirror is inside the cloud we observed strong deviations from the thin medium model even for the period of the patterns, which does not decrease to zero but saturates at a finite minimum value. Such a behavior can be expected as the medium thickness becomes the relevant diffractive length scale once the mirror distance is small. A complex asymmetric behavior including mode jumps was observed, which is qualitatively reproduced by the local thick medium model but whose quantitative understanding requires further investigations. In particular, the inclusion of wavenumber dependent losses due to the residual atomic motion is expected to improve the theoretical description. 

These investigations provide an insight in the transition out of the Raman-Nath regime for a nonlinear grating. They might be also important to understand the limits of storing images and structured quantum information in atomic samples. 

\section{Appendix A: Quasi-Kerr thick medium threshold condition}
\label{appendix:A}
\setcounter{equation}{0}
\renewcommand{\theequation}{A\arabic{equation}}
\renewcommand{\theHsection}{A\arabic{section}}

We here recount the derivation of the threshold curve for the quasi-Kerr case of the thick medium model, given in \cite{Firth2017}, following the framework of linear stability analysis described in \cite{Firth1990b}. We start with the paraxial wave equations for the forward field $F$ and backward field $B$ inside the quasi-Kerr nonlinear medium:
\begin{align}\label{eq:paraxwave1}
\begin{aligned}
\frac{\partial F}{\partial z}&-\frac{iL}{2k}\nabla_\perp F=-\frac{i\Delta\alpha_lL}{2}(1+|F|^2+|B|^2)F,\\
\frac{\partial B}{\partial z}&+\frac{iL}{2k}\nabla_\perp B=\frac{i\Delta\alpha_lL}{2}(1+|F|^2+|B|^2)B,
\end{aligned}
\end{align}
where electric field intensity is normalized to the detuned saturation intensity of the transition $I_{s,\Delta}=I_s/(1+\Delta^2)$, $z$ is in units of the medium length $L$, $\alpha_l=\alpha_0/(1+\Delta^2)$, with $\mbox{OD}=\alpha_0 L$ being the medium's optical thickness. Here, we have neglected the medium's absorption, and we call this approximation the quasi-Kerr limit, along with the reflection grating (i.e. $G=1+h=1$ in \cite{Firth2017}). The latter can be neglected because the nonlinearity is sufficiently slow such that this small-period longitudinal grating is washed out by atomic motion during the pattern formation dynamics.

For the homogeneous case the solutions are:
\begin{align}\label{eq:homsol}
\begin{aligned}
F_0(z)&=F_0\exp\left[-\frac{i\Delta\alpha_lL}{2}(1+|F_0|^2+|B_0|^2)z\right],\\
B_0(z)&=B_0\exp\left[\frac{i\Delta\alpha_lL}{2}(1+|F_0|^2+|B_0|^2)z\right].
\end{aligned}
\end{align}
Writing $F(x,z)=F_0[1+f\cos(qx)]$, $B(x,z)=B_0[1+b\cos(qx)]$ leads to equations for the transverse perturbation functions $f,\:b$:
\begin{align}\label{eq:transversepert}
\begin{aligned}
\frac{\partial f}{\partial z}&=-i\theta f+i\Delta\alpha_l L(|F_0|^2f'+|B_0|^2b'),\\
\frac{\partial b}{\partial z}&=i\theta f-i\Delta\alpha_l L(|F_0|^2f'+|B_0|^2b'),
\end{aligned}
\end{align}
where $f=f'+if''$ and $b=b'+ib''$ and $\theta=q^2L/2k$ is the diffractive phase shift between $f$ and $F_0$ after traversing the length of the optical medium. The full system of equations can be written in matrix form for vector $\mathbf{r}=(f',b',f'',-b'')^\intercal$:
\begin{align}\label{eq:transversepert_matrixform}
\begin{aligned}
\frac{\partial\mathbf{r}}{\partial z}=\theta\hat{M}\mathbf{r},
\end{aligned}
\end{align}
where the matrix $\hat{M}$ has the form:
\begin{align}\label{eq:matrixm}
\begin{aligned}
\hat{M}=\begin{pmatrix}
0 & \mathbf{I}_2\\
-\hat{C} & 0 
\end{pmatrix},\:\mbox{where:} \:\:\hat{C}=\begin{pmatrix}
1-\varphi & -\beta\\
-\varphi & 1-\beta 
\end{pmatrix},
\end{aligned}
\end{align}
with $\beta =\Delta\alpha_lL|B_0|^2/\theta$ and $\varphi=\Delta\alpha_lL|F_0|^2/\theta$. In the following we put $|F_0|^2=|B_0|^2=I_0/I_{s,\Delta}$, leading to $\varphi=\beta=\Delta\alpha_lL|F_0|^2/\theta$. Equation (\ref{eq:transversepert_matrixform}) can be integrated to give $\mathbf{r}(z)=\exp(\theta\hat{M}z)\mathbf{r}(0)$, where the exponential of $\hat{M}$ at $z=1$ (end of the optical medium) can be readily calculated to give
\begin{align}\label{eq:expmatrixm}
\begin{aligned}
\exp(\theta\hat{M})=\begin{pmatrix}
\cos\theta\hat{c} & \hat{c}^{-1}\sin\theta\hat{c}\\
-\hat{c}\sin\theta\hat{c} & \cos\theta\hat{c} 
\end{pmatrix},
\end{aligned}
\end{align}
where $\hat{c}^2=\hat{C}$. Note that even though $\hat{c}$ is not a uniquely defined matrix, the elements of $\exp[\theta\hat{M}]$ are even functions of $\hat{c}$, i.e. they are uniquely defined matrices. Any analytical function $g(\hat{C})$ of $\hat{C}$ can be written using Sylvester's formula for $2\times 2$ matrices, leading to:
\begin{align}\label{eq:gofC}
\begin{aligned}
g(\hat{C})=\begin{pmatrix}
0 & -\varphi\\
-\varphi & 0 
\end{pmatrix}\frac{[g(\alpha_1)-g(\alpha_2)]}{\alpha_1-\alpha_2}+\frac{\mathbf{I}_2}{2}[g(\alpha_1)+g(\alpha_2)],
\end{aligned}
\end{align}
where $\alpha_j$ are eigenvalues of $\hat{C}$, given by $\alpha_{1,2}=(1,1-2\varphi)$. This expression can be used to evaluate $\exp(\theta\hat{M})$. Boundary conditions for the single-mirror feedback system are $f(0)=0,\: b(1)=e^{-2i\psi_D}f(1)$, where $\psi_D=D\theta$, and $D=d/L-1/2$ is the normalized mirror distance from the end of the optical medium. Implementing the boundary conditions via $\mathbf{r}(1)=\exp(\theta\hat{M})\mathbf{r}(0)$, leads to a system of equations:
\begin{widetext}
\begin{align}\label{eq:initial_eqs}
\begin{aligned}
f'(1)&=(\cos\theta\hat{c})_{12}b'(0)-(\hat{c}^{-1}\sin\theta\hat{c})_{12}b''(0),\\
f'(1)\cos 2\psi_D+f''(1)\sin 2\psi_D&=(\cos\theta\hat{c})_{22}b'(0)-(\hat{c}^{-1}\sin\theta\hat{c})_{22}b''(0),\\
f''(1)&=-(\hat{c}\sin\theta\hat{c})_{12}b'(0)-(\cos\theta\hat{c})_{12}b''(0),\\
f'(1)\sin 2\psi_D-f''(1)\cos 2\psi_D&=-(\hat{c}\sin\theta\hat{c})_{22}b'(0)-(\cos\theta\hat{c})_{22}b''(0).
\end{aligned}
\end{align}
\end{widetext}
After some algebra, these equations lead to the single-mirror feedback thick medium threshold condition \cite{Firth2017}:
\begin{align}\label{eq:threshold}
\begin{aligned}
c_1c_2+\left(\frac{\psi_2}{\psi_1}c_D^2+\frac{\psi_1}{\psi_2}s_D^2\right)s_1s_2=c_Ds_D(\beta_1s_1c_2-\beta_2s_2c_1),
\end{aligned}
\end{align}
where $\psi_1=\theta$, $\psi_2=\sqrt{\theta(\theta+2\Delta\alpha_lLI_0/I_{s,\Delta})}$ [from $\alpha_i=(\psi_i/\theta)^2$], $s_i=\sin\psi_i$, $c_i=\cos\psi_i$, $s_D=\sin\psi_D$, $c_D=\cos\psi_D$, $\beta_i=\left(\frac{\psi_i}{\theta}-\frac{\theta}{\psi_i}\right)$. Solving Eq. (\ref{eq:threshold}) leads to the threshold values of diffractive phase shift $\theta$ and nonlinear phase shift $\Delta\alpha_l LI_0/I_{s,\Delta}$. To compare with the experimental results, the diffractive phase shift is converted to the diffracted angle $\theta_d$ using $\theta_d=\sqrt{2\theta/kL}$.

\noindent
{\large \bf Acknowledgements}\\
{\footnotesize The Nice group acknowledges support from CNRS, UNS, and R\'{e}gion PACA. I. K. acknowledges support of the project KODYN financed by the European Union through the National Recovery and Resilience Plan 2021–2026, and the project Centre for Advanced Laser Techniques (CALT), cofunded by the European Union through the European Regional Development Fund under the Competitiveness and Cohesion Operational Programme (Grant No. KK.01.1.1.05.0001).

\end{document}